\documentclass[sigconf]{acmart}

\usepackage{graphicx}
\usepackage{subcaption}
\usepackage{balance}
\usepackage{hyperref} 
\usepackage{multirow}
\newcommand{\eat}[1]{}
\AtBeginDocument{%
  }

\copyrightyear{2025}
\acmYear{2025}
\setcopyright{acmlicensed}\acmConference[SIGIR '25]{Proceedings of the 48th International ACM SIGIR Conference on Research and Development in Information Retrieval}{July 13--18, 2025}{Padua, Italy}
\acmBooktitle{Proceedings of the 48th International ACM SIGIR Conference on Research and Development in Information Retrieval (SIGIR '25), July 13--18, 2025, Padua, Italy}
\acmDOI{10.1145/3726302.3729994}
\acmISBN{979-8-4007-1592-1/2025/07}

\begin{document}

\title{Hierarchical Intent-guided Optimization with Pluggable LLM-Driven Semantics for Session-based Recommendation}

\author{Jinpeng Chen}
\email{jpchen@bupt.edu.cn}
\affiliation{%
  \institution{School of Computer Science (National Pilot Software Engineering School), Beijing University of Posts and Telecommunications}
  \city{Beijing}
  \country{China}
}

\authornote{{Corresponding author.}}

\authornote{Also with Key Laboratory of Trustworthy Distributed Computing and Service (BUPT), Ministry of Education.}

\author{Jianxiang He}
\email{hjx812143280@bupt.edu.cn}
\affiliation{%
  \institution{School of Computer Science (National Pilot Software Engineering School), Beijing University of Posts and Telecommunications}
  \city{Beijing}
  \country{China}
}

\authornotemark[2]

\author{Huan Li}
\email{lihuan.cs@zju.edu.cn}
\affiliation{%
  \institution{The State Key Laboratory of Blockchain and Data Security, Zhejiang University}
  \city{Hangzhou}
  \country{China}
}

\author{Senzhang Wang}
\email{szwang@csu.edu.cn}
\affiliation{%
  \institution{Central South University}
  \city{Changsha}
  \country{China}
}

\author{Yuan Cao}
\email{e1124923@u.nus.edu}
\affiliation{%
  \institution{National University of Singapore}
  \city{Singapore}
  \country{Singapore}
}
\author{Kaimin Wei}
\email{cswei@jnu.edu.cn}
\affiliation{%
  \institution{Jinan University}
  \city{Guangzhou}
  \country{China}
}

\author{Zhenye Yang}
\email{yzy@bupt.edu.cn}
\affiliation{%
  \institution{Beijing University of Posts and Telecommunications}
  \city{Beijing}
  \country{China}
}

\author{Ye Ji}
\email{jiye@travelsky.com.cn}
\affiliation{%
  \institution{TravelSky Technology Limited}
  \city{Beijing}
  \country{China}
}

\renewcommand{\shortauthors}{Jinpeng Chen et al.}

\begin{abstract}

Session-based Recommendation (SBR) aims to predict the next item a user will likely engage with, using their interaction sequence within an anonymous session. Existing SBR models often focus only on single-session information, ignoring inter-session relationships and valuable cross-session insights. Some methods try to include inter-session data but struggle with noise and irrelevant information, reducing performance. Additionally, most models rely on item ID co-occurrence and overlook rich semantic details, limiting their ability to capture fine-grained item features. To address these challenges, we propose a novel hierarchical intent-guided optimization approach with pluggable LLM-driven semantic learning for session-based recommendations, called HIPHOP. First, we introduce a pluggable embedding module based on large language models (LLMs) to generate high-quality semantic representations, enhancing item embeddings. Second, HIPHOP utilizes graph neural networks (GNNs) to model item transition relationships and incorporates a dynamic multi-intent capturing module to address users' diverse interests within a session. Additionally, we design a hierarchical inter-session similarity learning module, guided by user intent, to capture global and local session relationships, effectively exploring users' long-term and short-term interests. To mitigate noise, an intent-guided denoising strategy is applied during inter-session learning. Finally, we enhance the model's discriminative capability by using contrastive learning to optimize session representations. Experiments on multiple datasets show that HIPHOP significantly outperforms existing methods, demonstrating its effectiveness in improving recommendation quality. Our code is available: \url{https://github.com/hjx159/HIPHOP}.

\end{abstract}




\maketitle

\section{Introduction}

Recommendation systems (RS) are essential for navigating vast content and reducing information overload. Traditional methods like matrix factorization \cite{DBLP:journals/computer/KorenBV09} and collaborative filtering \cite{DBLP:conf/www/SarwarKKR01} rely on user profiles and extensive historical data but struggle with new or anonymous users due to limited data and privacy issues \cite{latifi2021session, guo2019streaming}, and often fail to capture the dynamic nature of user interests \cite{wang2021survey}. Session-based recommendation (SBR) \cite{li2017neural,DBLP:journals/corr/HidasiKBT15} addresses these issues by predicting the next actions from short, anonymous interaction sequences without relying on user identities. This makes SBR particularly valuable in real-time environments such as e-commerce and video sharing.

\begin{figure}[htbp]
  \centering
  
  \begin{subfigure}[t]{\linewidth}
    \centering
    \includegraphics[width=\linewidth]{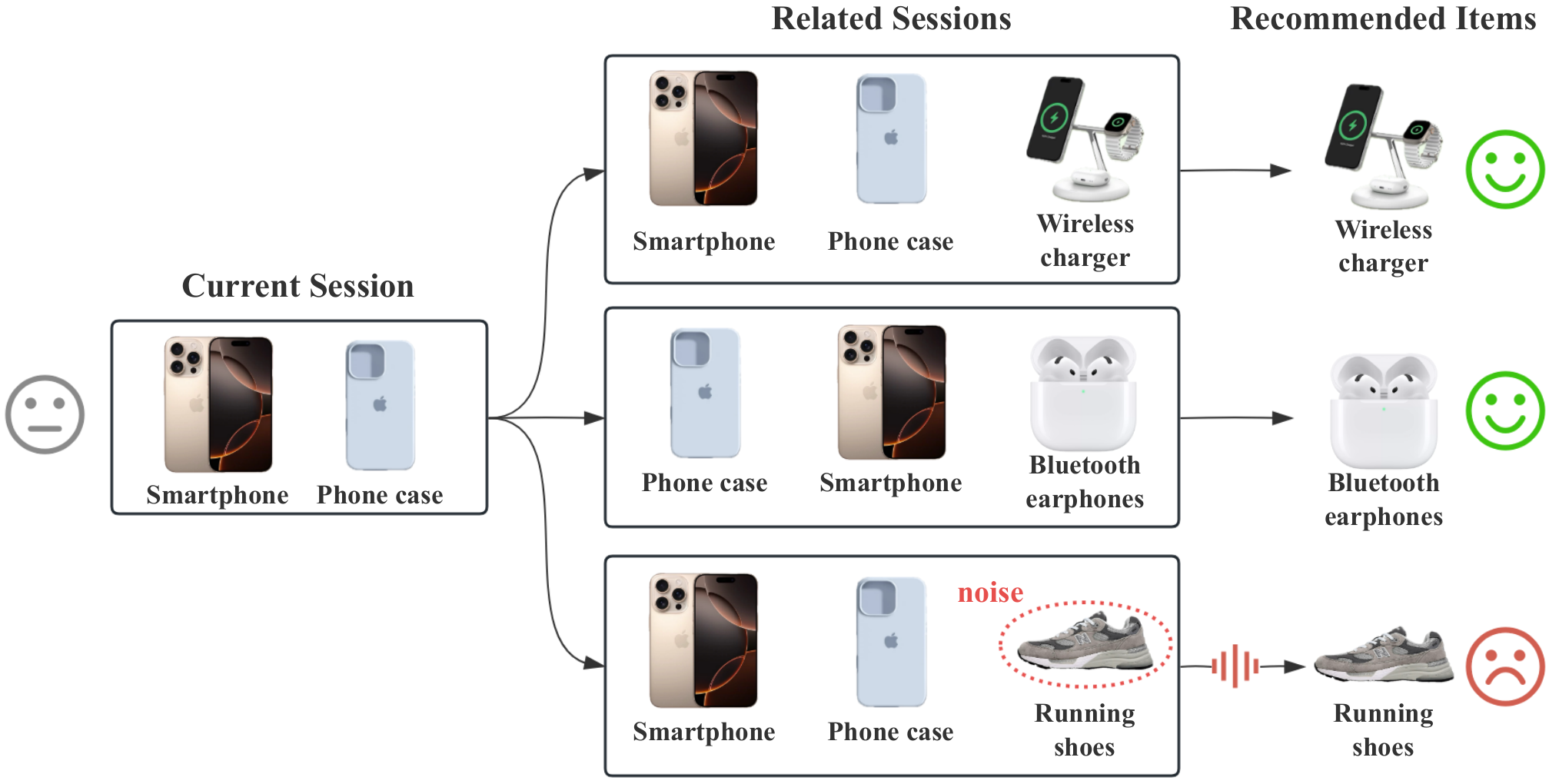}
    \subcaption{Session Correlations and Noise}
    \label{fig:fig1a}
  \end{subfigure}
  
  \begin{subfigure}[t]{\linewidth}
    \centering
    \includegraphics[width=\linewidth]{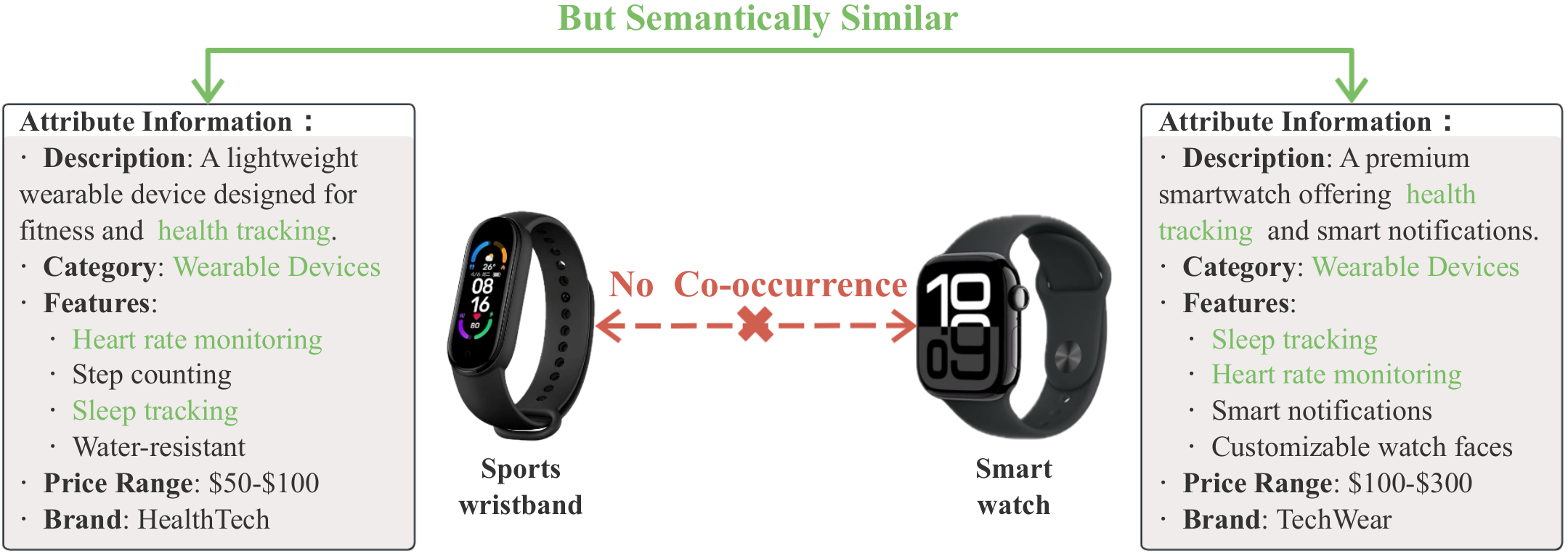}
    \subcaption{Item ID Co-occurrence Flaws}
    \label{fig:fig1b}
  \end{subfigure}
  
  \caption{An example of the limitations of current SBR models in inter-session associations and semantics usage.}
  \label{fig:fig1}
\end{figure}


Early SBR methods include pattern and rule mining \cite{adomavicius2005toward, wang2017inferring, shao2009music} and Markov chain-based approaches \cite{DBLP:journals/jmlr/ShaniHB05}. While rule mining efficiently extracts session-based associations, it often overlooks sequential patterns. Markov chain methods model user behavior sequences but assume high action independence and primarily focus only on recent interactions, limiting their ability to capture broader contextual understanding.

With the rise of deep learning, advanced SBR methods using recurrent neural networks (RNN) \cite{DBLP:journals/corr/HidasiKBT15, li2017neural} and attention mechanisms \cite{liu2018stamp} emerged, effectively capturing temporal sequences but still lacking in depicting complex item relationships. Graph neural network (GNN)-based methods like SR-GNN \cite{wu2019session} have become mainstream and are effective in modeling complex item transitions \cite{DBLP:conf/ijcai/XuZLSXZFZ19, yu2020tagnn, guo2022evolutionary, qiu2019rethinking, chen2023attribute}. However, these approaches mainly focus on current sessions, neglecting inter-session associations and valuable cross-session patterns. For example, as shown in Figure \ref{fig:fig1a}, traditional methods might recommend additional phone models based on current browsing "smartphone" and "phone case," while cross-session information could suggest complementary accessories such as "wireless chargers" or "Bluetooth earphones".

Researchers have attempted to integrate collaborative information from neighboring sessions \cite{wang2019collaborative} or to construct global session graphs \cite{wang2020global} to leverage cross-session item transitions. However, these methods often handle information from a single perspective and are susceptible to noisy data, which can introduce irrelevant items and degrade recommendation quality. As demonstrated in Figure \ref{fig:fig1a}, for example, adding "running shoes" to a session browsing "smartphone" can mislead the system.

Additionally, most existing SBR methods rely primarily on user interaction data, learning the item representations based on co-occurrence patterns between item IDs \cite{DBLP:journals/tkde/ZhangXMLYL24}. This approach lacks a semantic understanding of user-item interactions, such as titles, descriptions, and attributes, limiting the ability to capture detailed item features and reducing recommendation accuracy. As shown in Figure \ref{fig:fig1b}, “sports wristband” and “smartwatch” may rarely co-occur despite their semantic similarity, resulting in missed opportunities for personalized recommendations.

To address these challenges, we propose a novel SBR model called HIPHOP. This model incorporates semantic embeddings from LLMs to enhance item representations. It combines GNN and hierarchical cross-session similarity learning to effectively capture complex intra-session transitions and multi-level inter-session associations. Additionally, HIPHOP models multiple user intents to capture diverse interests within the current session, reducing noise in cross-session learning. Finally, introducing contrastive learning to optimize session representation has improved discriminative ability and recommendation accuracy.


Our main contributions are as follows:
\begin{itemize}
    \item We introduce HIPHOP, an SBR method that combines dynamic multi-intent capture, hierarchical inter-session similarity, contrastive learning, and a pluggable LLM-driven semantic embedding module. This pluggable module is compatible with most existing SBR models, enabling seamless integration of semantic information.
    \item We construct three novel SBR datasets with item semantic information. Unlike existing SBR datasets that rely solely on item ID co-occurrence, our datasets include detailed item attributes, providing the resource for advancing SBR tasks.
    \item We conduct extensive experiments on five datasets, demonstrating that HIPHOP outperforms baseline methods and achieves state-of-the-art performance.
\end{itemize}

\section{Related Work}

\subsection{Session-based Recommendations}


\subsubsection{Traditional Methods}
Early methods fall into two categories: (1) Pattern mining methods (S-POP~\cite{adomavicius2005toward}, IRRMiner~\cite{wang2017inferring}, DWA~\cite{shao2009music}) that extract item co-occurrence rules but neglect sequential dynamics; (2) Markov chain models (FPMC~\cite{rendle2010factorizing}, FPMC-LR~\cite{DBLP:conf/ijcai/ChengYLK13}, Fossil~\cite{he2016fusing}) that focus on immediate transitions while assuming action independence. Both categories struggle with long-term dependencies and complex behavioral patterns due to their localized modeling perspective.

\subsubsection{Deep Learning-Based Methods}
Deep learning has advanced SBR through models based on recurrent neural networks (RNNs) and attention mechanisms. GRU4Rec~\cite{DBLP:journals/corr/HidasiKBT15} and its variants~\cite{tan2016improved} leveraged GRUs to capture long-term dependencies, while NARM~\cite{li2017neural} and STAMP~\cite{liu2018stamp} utilized attention mechanisms to model sequential behaviors and session interests. Transformer-based models like SASRec~\cite{kang2018self}, ISLF~\cite{DBLP:conf/ijcai/Song0OZXL19}, and MCPRN~\cite{DBLP:conf/ijcai/Wang0WSOC19} further improved the ability to capture dynamic user preferences. Despite these advancements, these methods primarily focus on adjacent item dependencies and struggle with capturing complex transition patterns.

Graph Neural Network (GNN)-based methods have become mainstream in SBR due to their ability to model complex item transitions. SR-GNN~\cite{wu2019session} introduced session graphs and used GNNs to generate high-quality item embeddings. Subsequent models, such as GC-SAN~\cite{DBLP:conf/ijcai/XuZLSXZFZ19}, TAGNN~\cite{yu2020tagnn}, FGNN~\cite{qiu2019rethinking}, ADRL~\cite{chen2023attribute}, Atten-Mixer~\cite{zhang2023efficiently}, and HearInt~\cite{wang2024spatial}, have further enhanced GNN-based SBR by incorporating attention mechanisms, attribute information, and intent modeling. However, these approaches often focus solely on intra-session information, neglecting inter-session correlations that could enhance recommendation accuracy.

To integrate inter-session information, methods like CSRM~\cite{wang2019collaborative}, GCE-GNN~\cite{wang2020global}, HG-GNN~\cite{pang2022heterogeneous}, and HADCG~\cite{su2023enhancing} have been proposed. These models leverage collaborative information from neighboring sessions and construct global session graphs to capture cross-session item transitions. Despite improving performance, they often have a limited processing perspective and insufficient noise handling. For example, noise information such as "Running Shoes" as shown in Figure \ref{fig:fig1a} may appear, which may degrade recommendation quality. In contrast, HIPHOP not only utilizes GNNs for in-session transitions but also employs hierarchical inter-session similarity learning with intent-guided noise reduction, effectively capturing multi-level session correlations and mitigating noise.

\subsection{Contrastive Learning for Recommendation}

Contrastive learning has gained prominence in RS for enhancing model discriminative capabilities through sample comparison. Techniques like S3-Rec~\cite{zhou2020s3}, MCLRec~\cite{qin2023meta}, VGCL~\cite{yang2023generative}, and RealHNS~\cite{ma2023exploring} have applied contrastive learning to sequential and cross-domain recommendations, utilizing strategies such as data augmentation, meta-learning, and hard-negative sampling to improve performance. In SBR, methods like DHCN~\cite{xia2021self1}, COTREC~\cite{xia2021self2}, RESTC~\cite{wan2023spatio}, and STGCR~\cite{wang2023cross} have employed contrastive learning to optimize session representations by maximizing mutual information and capturing temporal dynamics. 
Different from existing works, this paper introduces a novel positive and negative sample sampling method to enhance session representations, thereby improving both recommendation accuracy and model robustness.

\subsection{LLM in Recommendation}


\subsubsection{LLM-based Recommendation Models}

LLMs can be used directly as recommendation models through prompt-based and fine-tuned instruction-based methods. Prompt-based approaches \cite{dai2023uncovering} use natural language instructions to generate recommendations, often enhanced by contextual content \cite{hou2024large}. However, they may underperform compared to traditional models due to the complexity of user-item interactions. Fine-tuning methods \cite{zheng2024adapting} adapt LLMs for recommendation tasks by training them on instructional data, including textual descriptions or index ID representations. Textual methods integrate item descriptions and user interactions into text-based instructions, while index ID methods use sequences of unique item IDs. Achieving semantic alignment between LLMs and collaborative semantics is crucial for optimal performance \cite{zheng2024adapting}.
However, due to the limitations of these prompt-based and fine-tuning approaches in effectively capturing complex user-item interactions, this paper does not adopt this strategy.

\subsubsection{LLM-Enhanced Recommendation Models}

LLMs also enhance RS by improving data input, semantic representation, and preference modeling. In the data input stage, LLMs can enrich user and item features by extracting detailed information from interaction histories and item descriptions~\cite{xi2024towards}. During the encoding stage, LLMs generate semantic representations for users and items, providing knowledge-rich input features that enhance performance~\cite{DBLP:journals/corr/abs-2305-11700}. Furthermore, LLMs can be jointly trained with traditional RS models to align preference representations, improving recommendation quality while reducing computational overhead during deployment~\cite{DBLP:journals/corr/abs-2306-02841}. In this paper, we utilize LLMs to generate high-quality item semantic embeddings from item metadata. These embeddings are integrated into our SBR model, enriching the semantic depth of item representations and enhancing accuracy.

\section{Preliminaries}


\subsection{Problem Definition}

Let \( V = \{ v_1, v_2, \dots, v_n \} \) denote the set of items. An anonymous session is represented as an ordered sequence of item interactions \( S = \{ v_1, v_2, \dots, v_l \} \), where \( v_i \in V \) denotes the \( i \)-th clicked item and \( l \) is the session length. The goal of SBR is to predict the next item \( v_{l+1} \in V \) that the user is most likely to click.

\subsection{Multi-Level Session Graph Structures}



Building upon the set of items \( V \) and the sessions \( S \) defined in the problem definition, we construct three types of graphs to effectively capture both intra-session item transitions and inter-session similarities: the session graph \( G_s \), the global session similarity graph \( G_g \), and the local session similarity graph \( G_l \). The session graph \( G_s \) follows the methodology of SR-GNN \cite{wu2019session} to model item transitions. While \( G_g \) and \( G_l \) are novel contributions of this paper that capture similarities between different sessions at different levels.

\subsubsection{Session Graph \( G_s \)}

The session graph \( G_s = (V_s, E_s) \) is a directed graph representing item transitions within a single session \( S \). Nodes \( V_s \) include all items in \( S \), and edges \( E_s \) connect consecutive items. Each edge \( (v_i, v_j) \) is assigned a weight that increments with each occurrence of the transition and is normalized by the sum of incoming weights for each node. 

\subsubsection{Global Session Similarity Graph \( G_g \)}

The global session similarity graph \( G_g = ( \mathcal{S}, E_g ) \) is an undirected graph where each node represents a session in the set \( \mathcal{S} = \{ S_1, S_2, \dots, S_m \} \). Edges \( E_g \) connect every pair of distinct sessions, with weights determined by the Jaccard similarity of their item sets. This similarity measures the overlap in user interactions between sessions. The degree matrix \( \mathbf{D}_g \) normalizes these weights by the sum of similarities for each session, capturing long-term interest similarities across the dataset.

\subsubsection{Local Session Similarity Graph \( G_l \)}

Similarly, the local session similarity graph \( G_l = ( \mathcal{S}, E_l ) \) is an undirected graph with the same node set as \( G_g \). However, the edge weights \( W_l(S_a, S_b) \) are based on the Jaccard similarity of the last-\( k \) items in each session, emphasizing short-term interest similarities and capturing the user's immediate and recent behaviors. The degree matrix \( \mathbf{D}_l \) normalizes these weights in the same manner as \( \mathbf{D}_g \), ensuring that the total similarity for each session is appropriately scaled. This local similarity complements the global similarity by offering insights into the users' current interests.

\section{The Proposed Method}


\begin{figure*}[htbp]
    \centering
    \includegraphics[width=0.975\textwidth]{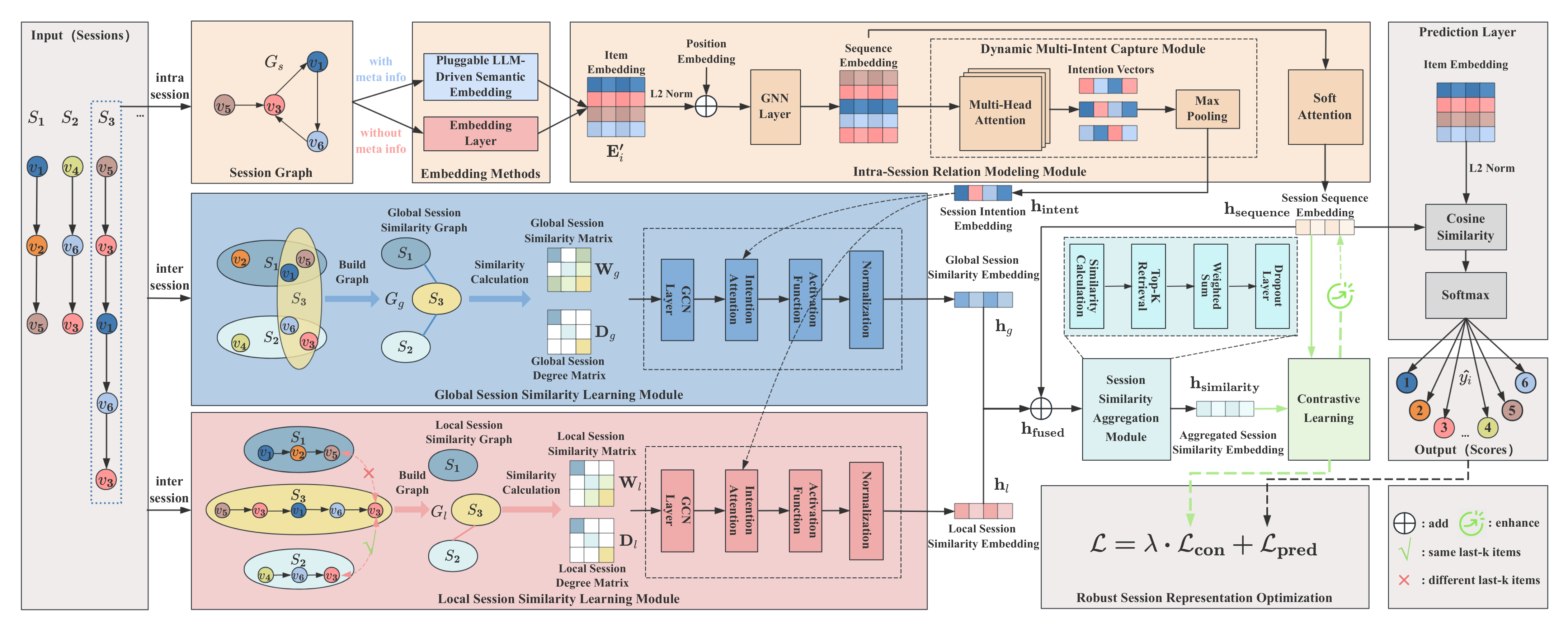}
    \caption{The architecture of \textit{HIPHOP} proposed.\eat{The model utilizes LLMs to generate semantically rich item embeddings and constructs a session graph, employing a GNN layer to capture item relationships within a session. A multi-head attention mechanism dynamically identifies multiple user intentions. It then learns hierarchical similarities between sessions and applies intent-guided attention to mitigate noise. Finally, session representations are optimized through contrastive learning to enhance recommendation accuracy.}}
    \label{fig:fig2}
\end{figure*}

As shown in Figure~\ref{fig:fig2}, the proposed HIPHOP starts with the \textit{Pluggable LLM-Driven Semantic Embedding Module} (cf. Section~\ref{sec:LLM-Driven Semantic Embedding Module}), which uses LLMs to generate semantically rich item embeddings, enhancing item representation with metadata. Next, the \textit{Intra-Session Relation Modeling Module} (cf. Section~\ref{sec:Intra-Session Relation Modeling Module}) constructs session graphs and applies GNNs to capture item transitions, followed by the \textit{Dynamic Multi-Intent Capture Module} (cf. Section~\ref{sec:Dynamic Multi-Intent Capture Module}), which employs multi-head attention to identify diverse user intentions from the session.
The \textit{Hierarchical Inter-Session Similarity Learning Module} (cf. Section~\ref{sec:Hierarchical Inter-Session Similarity Learning Module}) models both global and local inter-session similarities through global and local session similarity graph(cf. Section~\ref{sec:Global Session Similarity Learning Module} and Section~\ref{sec:Local Session Similarity Learning Module}), leveraging intent-guided attention to reduce noise. The \textit{Session Similarity Aggregation Module} (cf. Section~\ref{sec:Session Similarity Aggregation Module}) fuses these embeddings with intra-session representations to form aggregated session similarity embeddings. Finally, the \textit{Robust Session Representation Optimization Module} (cf. Section~\ref{sec:Robust Session Representation Optimization Module}) enhances session embeddings via contrastive learning (cf. Section~\ref{sec:Contrastive Learning}) and prediction optimization (cf. Section~\ref{sec:Prediction Layer}) , ensuring discriminative power and improved recommendation accuracy.

\subsection{LLM-Driven Semantic Embedding Module}\label{sec:LLM-Driven Semantic Embedding Module}

\begin{figure}[htbp]
    \centering
    \includegraphics[width=0.48\textwidth]{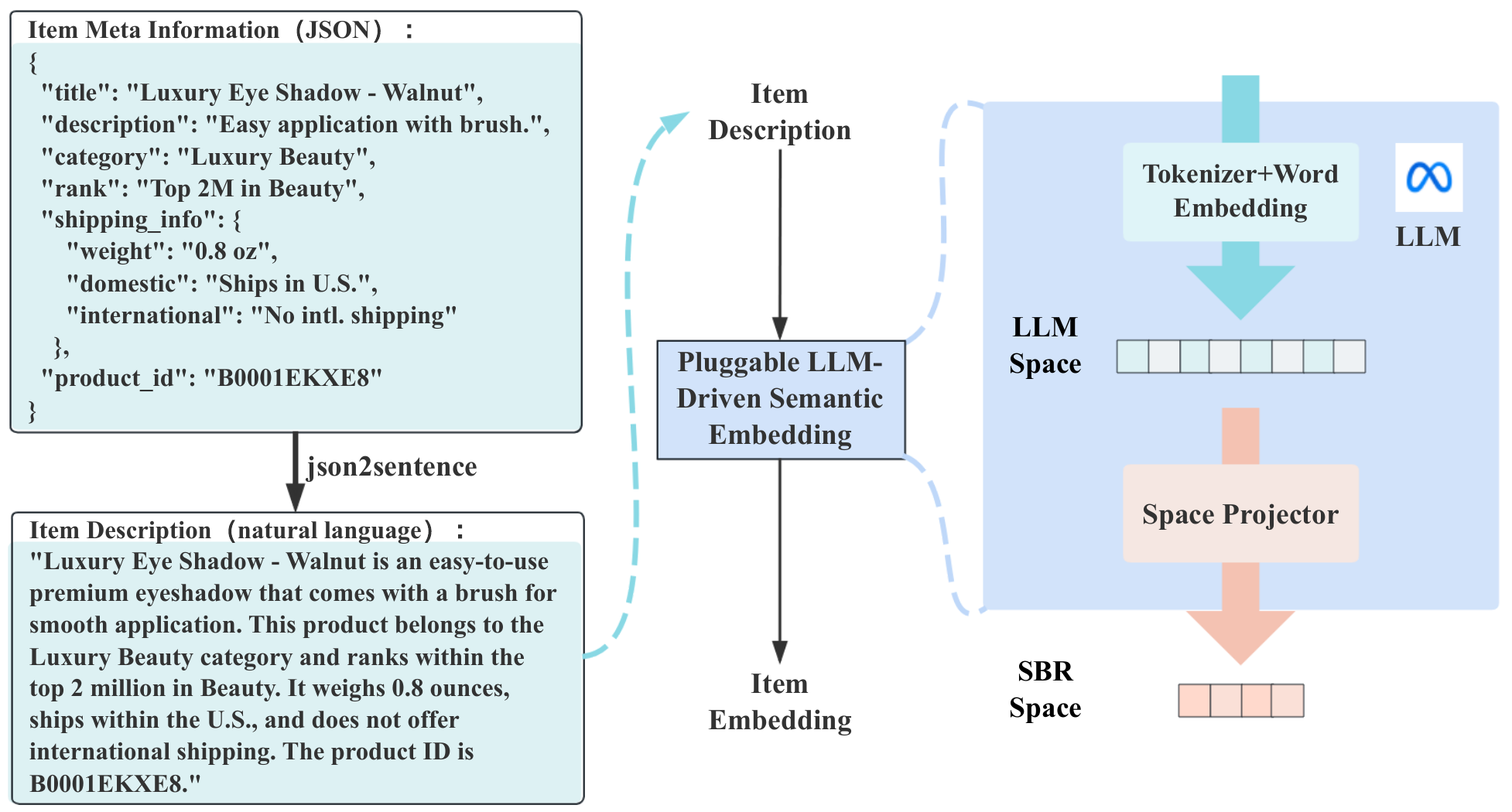}
    \caption{LLM-Driven Semantic Embedding Module.}
    \label{fig:fig3}
\end{figure}

As depicted in Figure ~\ref{fig:fig3}, this module enhances item representations by leveraging high-quality semantic embeddings generated by LLMs, thereby improving recommendation performance. Given that LLMs' reasoning abilities decline with standardized formats like JSON and XML~\cite{tam2024let}, we first convert item metadata into natural language descriptions using the \textit{json2sentence} method. These natural language descriptions are then input into the LLM to generate high-dimensional semantic embeddings \( \mathbf{E}_i \).

For items lacking metadata, their representations are initialized using the embedding layer. Subsequently, a \textit{Space Projector} submodule (e.g., a multilayer perceptron) maps the raw embeddings \( \mathbf{E}_i \) from the LLM embedding space to the hidden dimension \( d \) required by the SBR model, producing the mapped embedding \( \mathbf{E}'_i \).


This module enriches item representations with semantics, addressing the limitations of traditional SBR models that rely solely on item ID co-occurrence. Additionally, its pluggable design enables flexible integration or removal of components based on dataset characteristics, thereby enhancing the model's adaptability.

\subsection{Intra-Session Relation Modeling Module}\label{sec:Intra-Session Relation Modeling Module}

A session graph \( G_s = (V_s, E_s) \) is constructed, where \( V_s = \{ v_i \mid v_i \in S \} \) represents the items in session \( S \). The initial embedding \( \mathbf{h}^{(0)}_i \) for each item \( v_i \) is set to its semantic embedding if available. The GNN then updates the embeddings through multiple propagation steps:
\begin{equation}
\mathbf{h}^{(t+1)}_i = \sigma \left( \mathbf{W} \cdot \sum_{j \in \mathcal{N}(i)} w(v_j, v_i) \cdot \mathbf{h}^{(t)}_j \right)
\end{equation}
where \( \mathcal{N}(i) \) denotes the neighbors of item \( v_i \), \( w(v_j, v_i) \) is the normalized edge weight from \( v_j \) to \( v_i \), \( \mathbf{W} \) is a learnable weight matrix, and \( \sigma \) is an activation function, such as ReLU. After \( T \) propagation steps, the GNN produces updated item embeddings \( \mathbf{h}^{(T)} \) that capture high-order relationships.

Subsequently, a Soft Attention mechanism is applied to aggregate the updated item embeddings into a session representation \( \mathbf{h}_{\text{sequence}} \). This mechanism dynamically assigns weights to different items, allowing the model to capture the relative importance of each item and generate a comprehensive session representation that reflects the user's behavioral preferences:
\begin{equation}
\mathbf{h}_{\text{sequence}} = \sum_{i=1}^{l} \alpha_i \cdot \mathbf{h}^{(T)}_i
\end{equation}
where \( \alpha_i \) represents the attention weight for item \( v_i \), determined based on its relevance within the session. The Soft Attention mechanism allows the model to focus on more important items, thereby enhancing the quality of the session representation.

\subsection{Dynamic Multi-Intent Capture Module}\label{sec:Dynamic Multi-Intent Capture Module}

We initialize \( M \) learnable intent queries \( \mathbf{Q} = \{ \mathbf{q}_1, \mathbf{q}_2, \dots, \mathbf{q}_M \} \), where each \( \mathbf{q}_m \in \mathbb{R}^d \) represents a potential user intent. These queries interact with the item embeddings \( \mathbf{h}_i \) in the session to compute attention weights:
\begin{equation}
\alpha_{m,i} = \frac{\exp\left( \mathbf{q}_m^\top \cdot \mathbf{h}_i \right)}{\sum_{j=1}^{l} \exp\left( \mathbf{q}_m^\top \cdot \mathbf{h}_j \right)}
\end{equation}
where \( \alpha_{m,i} \) denotes the importance of item \( v_i \) for intent \( m \). Using these weights, we aggregate the item embeddings to form the intent-specific representation:
\begin{equation}
\mathbf{h}_{\text{intent}, m} = \sum_{i=1}^{l} \alpha_{m,i} \cdot \mathbf{h}_i
\end{equation}

Through the multi-head attention mechanism, each attention head focuses on different aspects of the session, such as functional characteristics, categories, or interaction order of items, allowing the model to capture diverse user intent patterns. The set of multiple intent representations \( \mathbf{H}_{\text{intent}} = \{ \mathbf{h}_{\text{intent}, 1}, \mathbf{h}_{\text{intent}, 2}, \dots, \mathbf{h}_{\text{intent}, M} \} \) is then aggregated using a \textit{Max Pooling} function to produce the final session intent representation:
\begin{equation}
\mathbf{h}_{\text{intent}} = \text{MaxPooling}\left(\mathbf{H}_{\text{intent}}\right)
\end{equation}
where the Max Pooling operation selects the most discriminative features from each intent vector, effectively capturing the core aspects of different user intents. This comprehensive session intent representation \( \mathbf{h}_{\text{intent}} \) is then utilized in subsequent modules to mitigate the impact of noise across sessions, thereby enhancing overall recommendation accuracy.

\subsection{Hierarchical Inter-Session Similarity Learning Module with Intent-Guided Noise Reduction}\label{sec:Hierarchical Inter-Session Similarity Learning Module}


\subsubsection{Global Session Similarity Learning Module}\label{sec:Global Session Similarity Learning Module}

We capture long-term session similarities using the global session similarity graph \( G_g = (\mathcal{S}, E_g) \). Given a session's sequence embedding \( \mathbf{h}_{\text{sequence}} = \{ \mathbf{h}_1, \mathbf{h}_2,\dots, \mathbf{h}_l \} \), we compute the initial global session representation \( \mathbf{h}_{\text{global}} = \sum_{i=1}^{l} \mathbf{h}_i \). We then apply graph convolution using the global similarity matrix \( \mathbf{W}_g \) and the degree matrix \( \mathbf{D}_g \) as follows:
\begin{equation}
\mathbf{h}_{\text{global}}' = \mathbf{D}_g \mathbf{W}_g \mathbf{h}_{\text{global}}
\end{equation}

To reduce noise, we apply an intent-guided attention mechanism:
\begin{equation}
\alpha_g = \text{softmax}\left( \text{ReLU}\left( \mathbf{W}_1 \mathbf{h}_{\text{global}}' + \mathbf{W}_2 \mathbf{h}_{\text{intent}} + b \right) \mathbf{W}_0^\top \right)
\end{equation}
where \( \mathbf{W}_1 \), \( \mathbf{W}_2 \) are learnable matrices, \( b \) is the bias vector, and \( \mathbf{W}_0 \) is the projection weight vector. The attention weight \( \alpha_g \) reflects the importance of each feature in the global similarity embedding, generating the denoised global similarity embedding:
\begin{equation}
\mathbf{h}_g = \alpha_g \cdot \mathbf{h}_{\text{global}}'
\end{equation}

\subsubsection{Local Session Similarity Learning Module}\label{sec:Local Session Similarity Learning Module}

Similarly, short-term session similarities are captured using the local session similarity graph \( G_l = (\mathcal{S}, E_l) \). The initial local session embedding is \( \mathbf{h}_{\text{local}} = \sum_{i=1}^{l} \mathbf{h}_i \). We then apply graph convolution using the local similarity matrix \( \mathbf{W}_l \) and the degree matrix \( \mathbf{D}_l \):
\begin{equation}
\mathbf{h}_{\text{local}}' = \mathbf{D}_l \mathbf{W}_l \mathbf{h}_{\text{local}}
\end{equation}

We apply the intent-guided attention mechanism as follows:
\begin{equation}
\alpha_l = \text{softmax}\left( \text{ReLU}\left( \mathbf{W}_1 \mathbf{h}_{\text{local}}' + \mathbf{W}_2 \mathbf{h}_{\text{intent}} + b \right) \mathbf{W}_0^\top \right)
\end{equation}

This results in the denoised local similarity embedding:
\begin{equation}
\mathbf{h}_l = \alpha_l \cdot \mathbf{h}_{\text{local}}'
\end{equation}

\subsubsection{Session Similarity Aggregation Module}\label{sec:Session Similarity Aggregation Module}

After hierarchical session similarity learning, we aggregate the global embedding \( \mathbf{h}_g \), the local embedding \( \mathbf{h}_l \), and the initial session representation \( \mathbf{h}_{\text{sequence}} \) by summing them to form the fused representation:
\begin{equation}
\mathbf{h}_{\text{fused}} = \mathbf{h}_{\text{sequence}} + \mathbf{h}_g + \mathbf{h}_l
\end{equation}

To enhance the quality of the session representation, we normalize the fused embeddings:
\begin{equation}
\tilde{\mathbf{h}}_{\text{fused}, i} = \frac{\mathbf{h}_{\text{fused}, i}}{\| \mathbf{h}_{\text{fused}, i} \|}, \quad \forall i \in \{1, 2, \dots, N\}
\end{equation}
where \( \| \mathbf{h}_{\text{fused}, i} \| \) denotes the Euclidean norm of \( \mathbf{h}_{\text{fused}, i} \). The normalized embeddings are then used to compute cosine similarities between sessions, generating the similarity matrix \( \text{CosSim} \).

For each session \( i \), we identify the top \( K \) most similar sessions based on \( \text{CosSim} \), forming the index set \( \text{TopK}_i \). The similarity weights are normalized using the Softmax function to obtain the contribution weights \( \alpha_{i,k} \). The similarity aggregated representation \( \mathbf{h}_{\text{similarity}, i} \) is then obtained by a weighted summation of the fused embeddings of the top \( K \) similar sessions:
\begin{equation}
\mathbf{h}_{\text{similarity}, i} = \sum_{k \in \text{TopK}_i} \alpha_{i,k} \cdot \mathbf{h}_{\text{fused}, k}
\end{equation}

Finally, a Dropout operation is applied to this representation to obtain the final similarity aggregated representation \( \mathbf{h}_{\text{similarity}} \).


\subsection{Robust Session Representation Optimization Module}\label{sec:Robust Session Representation Optimization Module}



\subsubsection{Contrastive Learning}\label{sec:Contrastive Learning}

We designate the current session's sequence representation \( \mathbf{h}_{\text{sequence}} \) as the anchor and the aggregated similarity representation \( \mathbf{h}_{\text{similarity}} \) as the positive sample. Negative samples are selected using Hard Negative Sampling, which chooses sessions similar to the current session but sharing no common items. The InfoNCE loss function is defined as:
\begin{equation}
\mathcal{L}_{\text{con}} = -\log \left( \frac{\exp\left( \text{sim}_{\text{pos}} / \tau \right)}{ \text{sim}_{\text{pos}} + \sum_{i=1}^{N_{\text{neg}}} \exp\left( \frac{\text{sim}_{\text{neg}, i}}{\tau} \right)} \right)
\end{equation}
where
\begin{equation}
\text{sim}_{\text{pos}} = \text{sim}\left( \mathbf{h}_{\text{sequence}}, \mathbf{h}_{\text{similarity}} \right)
\end{equation}
and
\begin{equation}
\text{sim}_{\text{neg}, i} = \text{sim}\left( \mathbf{h}_{\text{sequence}}, \mathbf{h}_{N_{\text{neg}, i}} \right)
\end{equation}
where \( \text{sim}(\cdot, \cdot) \) denotes cosine similarity, \( \mathbf{h}_{N_{\text{neg}, i}} \) is the embedding of the \( i \)-th negative sample, \( N_{\text{neg}} \) is the number of negative samples, and \( \tau \) is the temperature parameter. Additionally, \( \tau \) is dynamically adjusted during training to increase the difficulty of discrimination, thus promoting model stability and faster convergence.

\subsubsection{Prediction Layer}\label{sec:Prediction Layer}

We combine the sequence representation \eat{\( \mathbf{h}_{\text{sequence}} \)} and the similarity-aggregated embedding \eat{\( \mathbf{h}_{\text{similarity}} \)} to form the final session representation \( \mathbf{h}_{\text{session}} = \mathbf{h}_{\text{sequence}} + \mathbf{h}_{\text{similarity}} \). Next, we compute the prediction scores for each item \( v_j \) by measuring the similarity between \( \mathbf{h}_{\text{session}} \) and the item embeddings \( \mathbf{E}'_j \):
\begin{equation}
\text{score}(v_j) = \frac{ \mathbf{h}_{\text{session}}^\top \cdot \mathbf{E}'_j }{ \| \mathbf{h}_{\text{session}} \| \cdot \| \mathbf{E}'_j \| }
\end{equation}

To convert these scores into probabilities, we apply the softmax:
\begin{equation}
\hat{y}_{i,j} = \frac{\exp\left( \text{score}(v_j) \right)}{\sum_{k=1}^{n} \exp\left( \text{score}(v_k) \right)}
\end{equation}
where \( \hat{y}_{i,j} \) is the predicted probability for item \( v_j \) in session \( i \), and \( n \) is the total number of candidate items.

Prediction loss \( \mathcal{L}_{\text{pred}} \) is calculated using the cross entropy loss:
\begin{equation}
\mathcal{L}_{\text{pred}} = -\frac{1}{N} \sum_{i=1}^N \sum_{j=1}^n \left[ y_{i,j} \log( \hat{y}_{i,j} ) + (1 - y_{i,j}) \log( 1 - \hat{y}_{i,j} ) \right]
\end{equation}
where \( y_{i,j} \) is the ground-truth label indicating whether item \( v_j \) was clicked in session \( i \), and \( N \) is the number of training samples.

To optimize the model, we employ a joint loss function that combines the prediction loss \( \mathcal{L}_{\text{pred}} \) with the contrastive loss \( \mathcal{L}_{\text{con}} \):
\begin{equation}
\mathcal{L} = \mathcal{L}_{\text{pred}} + \lambda \cdot \mathcal{L}_{\text{con}}
\end{equation}
where \( \lambda \) is a hyperparameter that controls the weight of the contrastive loss in the total loss. This joint optimization strategy not only improves recommendation accuracy but also enhances the discriminative power of session representations, leading to more robust and reliable session-based recommendations.

\section{Experiments and Results}

We conducted experiments to validate HIPHOP's effectiveness by addressing the following questions:

\begin{itemize}
    \item \textbf{RQ1}: How does our model compare to state-of-the-art methods? (cf. Section~\ref{sec:RQ1})
    \item \textbf{RQ2}: Does each proposed technique improve model performance? (cf. Section~\ref{sec:RQ2})
    \item \textbf{RQ3}: How sensitive is the model to hyperparameter changes? (cf. Section~\ref{sec:RQ3})
    \item \textbf{RQ4}: What impact does the LLM-driven semantic embedding module have on recommendation performance? Can it improve other SBR models as well? (cf. Section~\ref{sec:RQ4})
\end{itemize}

\subsection{Experimental Settings}

\subsubsection{Datasets and Preprocessing}

To evaluate HIPHOP, we employ two public SBR datasets (Diginetica\footnotemark[1] and Yoochoose\footnotemark[2]) and three purpose-built Amazon\footnotemark[3]-derived datasets covering luxury beauty, musical instruments, and prime pantry categories. Unlike existing SBR datasets that only record item IDs, our Amazon variants additionally incorporate structured item attribute fields (titles, descriptions, category labels) to enable item semantic modeling.
Specifically, we preprocess these Amazon-derived datasets by treating each user review as an interaction and forming sessions by chronologically ordering reviews. Additionally, we convert item metadata into natural language descriptions and generate semantic embeddings using an LLM, thereby incorporating rich semantic information into the SBR model. Following the preprocessing steps outlined in~\cite{wu2019session,DBLP:conf/ijcai/XuZLSXZFZ19,wang2020global}, we filter out sessions with single interactions and remove items appearing fewer than five times. For each session \( S = [s_1, s_2, \dots, s_n] \), training and testing sequences were generated as \(([s_1], s_2), ([s_1, s_2], s_3), \dots, ([s_1, s_2, \dots, s_{n-1}], s_n)\). Table \ref{tab:dataset_statistics} presents the statistics of all five datasets.

\footnotetext[1]{\url{https://competitions.codalab.org/competitions/11161}}
\footnotetext[2]{\url{http://2015.recsyschallenge.com/challege.html}}
\footnotetext[3]{\url{https://nijianmo.github.io/amazon/index.html}}

\renewcommand{\arraystretch}{1.1}
\begin{table}[htbp]
\centering
\caption{Dataset Statistics}
\label{tab:dataset_statistics}
\resizebox{0.45\textwidth}{!}{
\begin{tabular}{l c c c c c}
\hline
Dataset             & Items   & Clicks  & Train   & Test    & Avg.len \\ \midrule
Diginetica                   & 43,097  & 982,961 & 719,470 & 60,858  & 5.12    \\ 
Yoochoose 1/64               & 16,766  & 557,248 & 369,859 & 55,898  & 6.16    \\ 
Luxury Beauty                & 1,438   & 33,864  & 3,213   & 603     & 8.87    \\ 
Musical Instruments          & 10,479  & 230,910 & 25,341  & 2,182   & 8.39    \\ 
Prime Pantry                 & 4,963   & 137,698 & 11,854  & 2,318   & 9.72    \\ \hline
\end{tabular}
}
\end{table}

\subsubsection{Evaluation Metrics}

We used two widely recognized evaluation metrics: \textbf{HR@K} (Hit Rate) and \textbf{MRR@K} (Mean Reciprocal Rank). 
HR@K measures whether the target item appears in the top K recommendations. MRR@K calculates the average reciprocal rank of the target item in the recommendation list, reflecting the model’s ability to rank the correct items higher. 
Similar to \cite{wu2019session}, we set \( K = 20 \) in this work.

\subsubsection{Baselines}

For a comprehensive comparison, we selected a diverse set of representative SBR models as baselines.

\renewcommand{\arraystretch}{1.0}
\begin{table}[h!]
\centering
\caption{Experimental Results on Diginetica and Yoochoose 1/64. The best method in each column is \textbf{boldfaced}, the second best is \underline{underlined}, and "–" indicates unavailable results in the original paper. Improv.(\%) denotes relative improvement between our method and the best baseline.}
\label{tab:overall_performance_1}
\begin{tabular}{l c c c c}
\hline
\multirow{2}{*}{\centering \text{Method}} & \multicolumn{2}{c}{{Diginetica}} & \multicolumn{2}{c}{{Yoochoose 1/64}} \\ 
{} & {HR@20} & {MRR@20} & {HR@20} & {MRR@20} \\ \midrule
{POP} & 1.18 & 0.28 & 4.51 & 0.72 \\ 
{S-POP} & 21.06 & 13.68 & 29.30 & 18.07 \\ 
{Item-KNN} & 35.75 & 11.57 & 52.13 & 21.44 \\ 
{FPMC} & 26.53 & 6.95 & 57.01 & 21.17  \\ \hline
{GRU4Rec} & 29.45 & 8.33 & 66.70 & 28.50  \\ 
{NARM} & 49.70 & 16.17 & 70.13 & 29.34  \\ 
{STAMP} & 45.64 & 14.32 & 68.74 & 29.67  \\ \hline
{SR-GNN} & 50.73 & 17.59 & 70.57 & 30.94   \\ 
{TAGNN} & 51.31 & 18.03 & 71.02 & 31.12  \\ \hline
{CSRM} & 50.55 & 16.38 & 71.45 & 30.36  \\ 
{GCE-GNN} & 54.22 & 19.04 & 70.91 & 30.63  \\ \hline
{COTREC} & 53.18 & 18.44 & 70.89 & 29.50  \\ \hline
{Atten-Mixer} & \underline{55.66} & 18.96 & \underline{72.51} & \underline{32.13} \\ 
{HearInt} & 55.02 & \underline{19.52} & - & -  \\ \hline
{HIPHOP} & \textbf{62.11} & \textbf{22.37} & \textbf{75.08} & \textbf{32.81} \\
{Improv.(\%)} & 11.59 & 14.60 & 3.48 & 1.46 \\ \hline
\end{tabular}
\end{table}

\begin{enumerate}

    \item \textbf{Traditional Methods}: \textbf{POP and S-POP}~\cite{adomavicius2005toward} recommend the top K most popular items overall and within the current session, respectively. \textbf{Item-KNN}~\cite{DBLP:conf/www/SarwarKKR01} recommends items similar to those previously clicked. \textbf{FPMC}~\cite{rendle2010factorizing} combines matrix factorization with Markov chains to capture preferences and patterns.


    \item \textbf{Sequence-based Models}: \textbf{GRU4Rec}~\cite{DBLP:journals/corr/HidasiKBT15} uses GRU with ranking loss for user sequences. \textbf{NARM}~\cite{li2017neural} adds an attention mechanism to GRU4Rec to capture user intent. \textbf{STAMP}~\cite{liu2018stamp} focuses on recent interests using short-term memory networks with self-attention.
    

    \item \textbf{GNN-based Models}: \textbf{SR-GNN}~\cite{wu2019session} uses GCNs on session graphs for item embeddings. \textbf{TAGNN}~\cite{yu2020tagnn} applies target-aware attention to model item transitions and user interests.
    

    \item \textbf{Inter-session Models}: \textbf{CSRM}~\cite{wang2019collaborative} combines RNN and attention with neighborhood session data. \textbf{GCE-GNN}~\cite{wang2020global} builds co-occurrence graphs to integrate local and global item information.
    

    \item \textbf{Contrastive Learning Models}: \textbf{COTREC}~\cite{xia2021self2} improves SBR through self-supervised and contrastive learning.
    

    \item \textbf{Multi-intent Models}: \textbf{Atten-Mixer}~\cite{zhang2023efficiently} models multi-granularity user intents. \textbf{HearInt}~\cite{wang2024spatial} enhances intent recognition with hierarchical spatio-temporal awareness and cross-scale contrastive learning.
\end{enumerate}

\subsubsection{Implementation Details}
To ensure fair comparisons with baselines, we followed the experimental setups in~\cite{wu2019session, wang2020global} and set the embedding dimension to 100. We utilized the Adam optimizer with an initial learning rate of 0.001, which decays by a factor of 0.1 every three epochs. An L2 regularization parameter of \(10^{-5}\) was applied to prevent overfitting.
An early stopping strategy was employed to halt training if no performance improvement was observed over three consecutive epochs. 
We adopt the embedding-3 model from Zhipu AI as the LLM to generate item semantic embeddings. Please note that this paper focuses not on comparing the performance of different LLMs but on introducing semantic information of LLM-driven items into the SBR task to enhance recommendation accuracy. The selection of embedding-3 is merely an attempt and serves as a reference, and readers are encouraged to experiment with other LLMs to evaluate the quality of semantic embeddings.
Our source code and preprocessed datasets are publicly available: \url{https://github.com/hjx159/HIPHOP}.

\eat{\footnotetext[4]{\url{https://github.com/xxx/xxx}}}

\subsection{Overall Performance (RQ1)}\label{sec:RQ1}
To further demonstrate the overall performance of our HIPHOP, we compare it with the selected baselines described above. The experimental results, presented in Tables \ref{tab:overall_performance_1} and \ref{tab:overall_performance_2}, cover two public datasets (Diginetica and Yoochoose 1/64), as well as three Amazon-derived datasets with item metadata (Luxury Beauty, Musical Instruments, and Prime Pantry). The results indicate that HIPHOP consistently outperforms all baseline models across all datasets.

\renewcommand{\arraystretch}{1.0}
\begin{table}[h!]
\caption{Experimental Results on Luxury Beauty, Musical Instruments, and Prime Pantry.}
\label{tab:overall_performance_2}
\resizebox{0.48\textwidth}{50pt}{
\begin{tabular}{l c c c c c c}
\hline
\multirow{2}{*}{\text{Method}} & \multicolumn{2}{c}{Luxury Beauty} & \multicolumn{2}{c}{Musical Instruments} & \multicolumn{2}{c}{Prime Pantry} \\ 
{} & HR@20 & MRR@20 & HR@20 & MRR@20 & HR@20 & MRR@20 \\ \midrule
{SR-GNN} & 30.65 & 18.40 & 22.03 & 11.90 & 17.37 & 6.14 \\ 
{TAGNN} & 30.81 & 18.42 & 22.28 & 11.67 & 17.16 & 5.96 \\ 
{GCE-GNN} & 30.73 & 17.54 & 19.44 & 9.36 & 14.93 & 4.16 \\ 
{COTREC} & 37.02 & 19.56 & 16.24 & 5.66 & 17.07 & 5.64 \\ 
{Atten-Mixer} & \underline{40.12} & \underline{23.28} & \underline{24.16} & \underline{12.90} & \underline{21.43} & \underline{9.16} \\ \hline
{HIPHOP} & \textbf{53.30} & \textbf{29.95} & \textbf{39.33} & \textbf{19.83} & \textbf{37.84} & \textbf{16.42} \\ 
{Improv.(\%)} & 32.85 & 28.65 & 62.79 & 53.72 & 76.57 & 79.26 \\ \hline
\end{tabular}
}
\end{table}

\begin{figure}[h]
    \centering
    \includegraphics[width=0.48\textwidth]{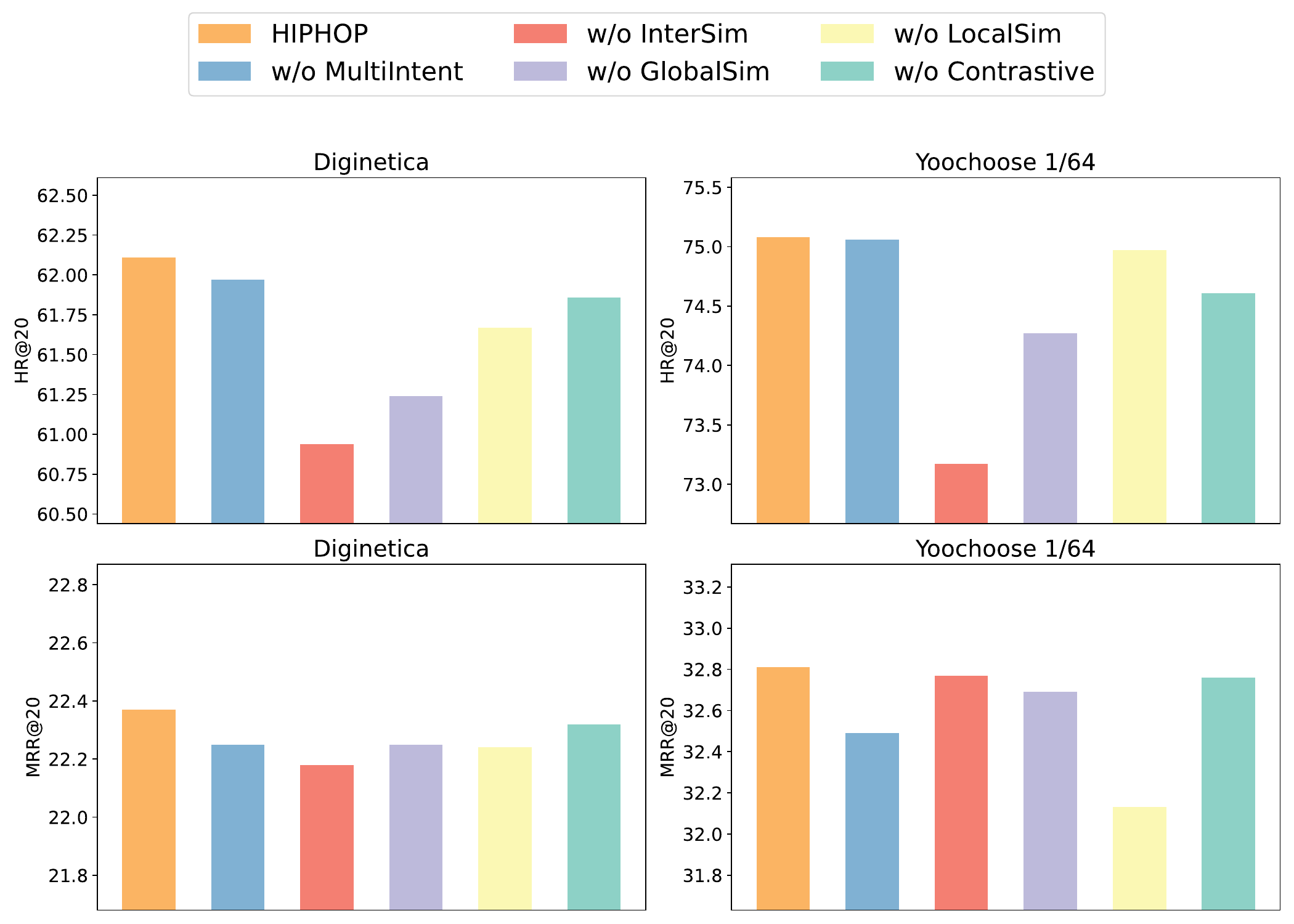}
    \caption{Ablation Study Results.}
    \label{fig:ablation_study_results}
\end{figure}

\begin{figure*}[h]
    \centering
    \includegraphics[width=0.90\textwidth]{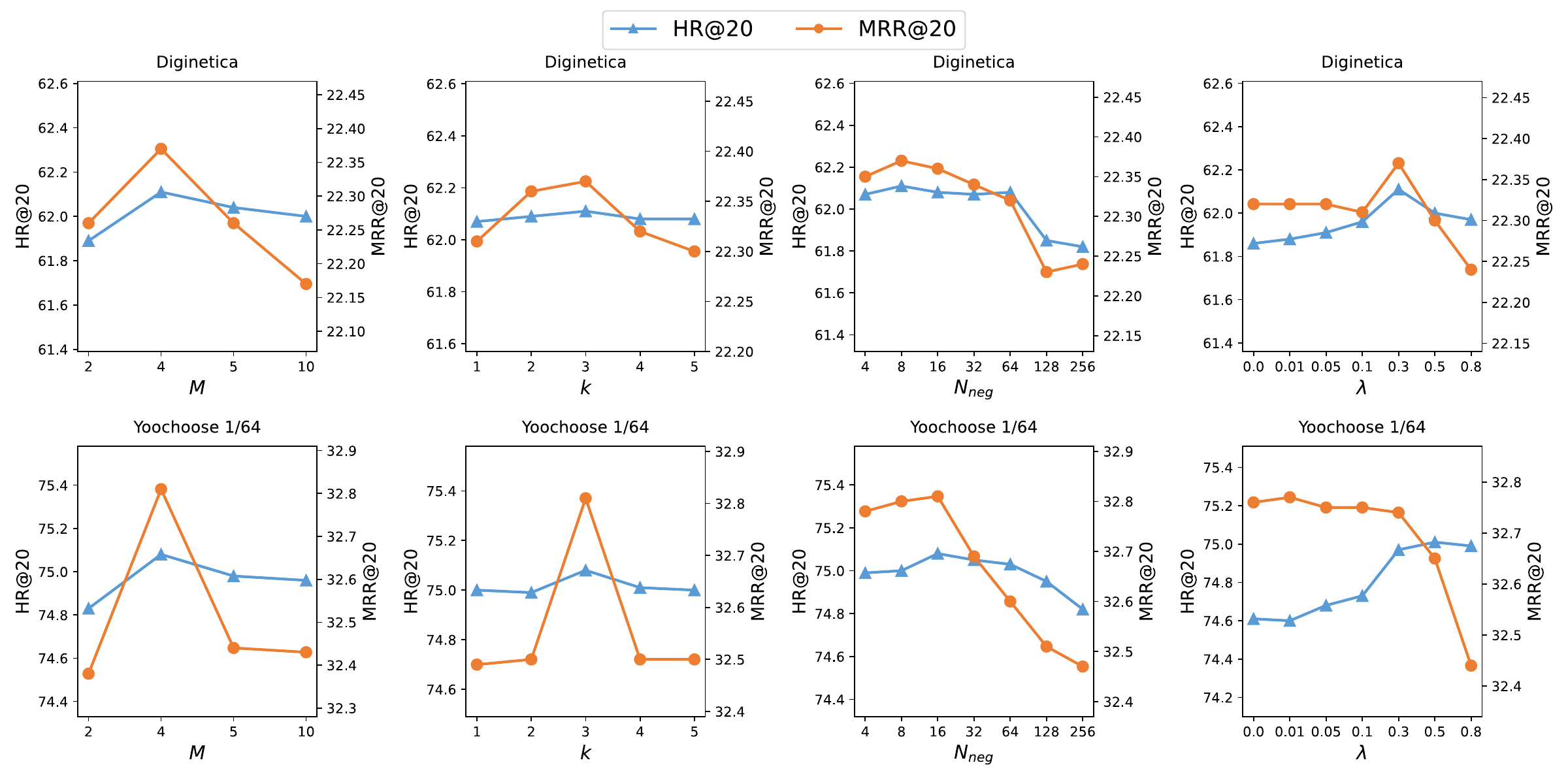}
    \caption{Impact of hyperparameters on HIPHOP's performance.}
    \label{fig:hyperparameters_study_results}
\end{figure*}

Among the traditional methods, POP and S-POP perform relatively poorly due to their simplistic strategies, which rely solely on item popularity and fail to leverage session-based information for modeling user behavior. FPMC which utilizes first-order Markov chains and matrix factorization, shows its effectiveness on two public datasets. Item-KNN achieves almost the best results among the traditional methods on the Diginetica and Yoochoose 1/64 datasets. However, it only applies the similarity between items and does not account for the chronological order of the items in a session, limiting its ability to capture sequential item transitions.

Compared with traditional methods, neural network-based methods generally perform better for SBR. Despite performing slightly worse than Item-KNN on Diginetica, GRU4Rec, as the first RNN-based method for SBR, demonstrates the capability of RNN to effectively model sequential data. NARM enhances the original RNN by incorporating an attention mechanism, which assigns different weights to items at various positions within the session. This results in a significant performance improvement over GRU4Rec, highlighting the effectiveness of attention mechanisms. STAMP, which replaces RNN with attentional MLPs, shows comparable performance over NARM. However, both RNNs and MLPs struggle to capture complex inter-session transitions, which may explain why they underperform compared to graph-based methods.

GNN-based models such as SR-GNN, and TAGNN significantly outperform the methods mentioned above by constructing session graphs that capture complex item transition relationships. Building on this, GCE-GNN achieves further performance gains by constructing additional global graphs, emphasizing the importance of inter-session data and proving the feasibility of creating additional graphs. Similarly, CSRM combines RNN and attention mechanisms with neighbor session data to better understand session intentions, achieving performance comparable to GNN-based methods and demonstrating the effectiveness of using item transitions from other sessions. In addition, CSRM treats other sessions as a whole, without distinguishing the relevant item-transitions from the irrelevant ones encoded in other sessions. COTREC, on the other hand,  employs effective data augmentation through self-supervised collaborative training, leveraging session graphs from dual perspectives, and utilizes contrastive learning to enhance the discriminative power of session representations. 

Advanced models that incorporate multi-intent modeling, such as Atten-Mixer and HearInt, deliver the best performance across nearly all selected baselines by effectively identifying diverse user intentions in session data. Notably, Atten-Mixer achieves the highest baseline performance, with substantial HR@20 scores on both Diginetica and Yoochoose 1/64 datasets, as well as MRR@20 scores on Yoochoose 1/64. These results highlight the effectiveness of multi-intent modeling in capturing complex user behavior.

As shown in Table \ref{tab:overall_performance_1}, HIPHOP has significantly improved performance compared to the best baseline, achieving the highest scores on all metrics on both public datasets. Specifically, On Diginetica and Yoochoose 1/64 datasets, HIPHOP achieved relative improvements of 11.59\% and 3.48\% respectively over the best baseline, Atten-Mixer, with HR@20 Scores of 62.11\% and 75.08\%. In addition, we extended the validation of HIPHOP effectiveness by incorporating three Amazon-derived datasets into the experiment. Table \ref{tab:overall_performance_2} shows the experimental results, in which HIPHOP exhibits significant performance, with relative improvements ranging from 28.65\% to 79.26\% on all evaluation metrics across the three datasets, significantly exceeding the optimal baseline Atten-Mixer. Our analysis of all five datasets shows that despite differences in data distribution and session length, HIPHOP consistently and significantly achieved promising results. This further confirms the effectiveness and superiority of HIPHOP in capturing user behavior patterns and utilizing semantic information, positioning it most advanced method in SBR.

\begin{figure*}[h]
    \centering
    \includegraphics[width=0.91\textwidth]{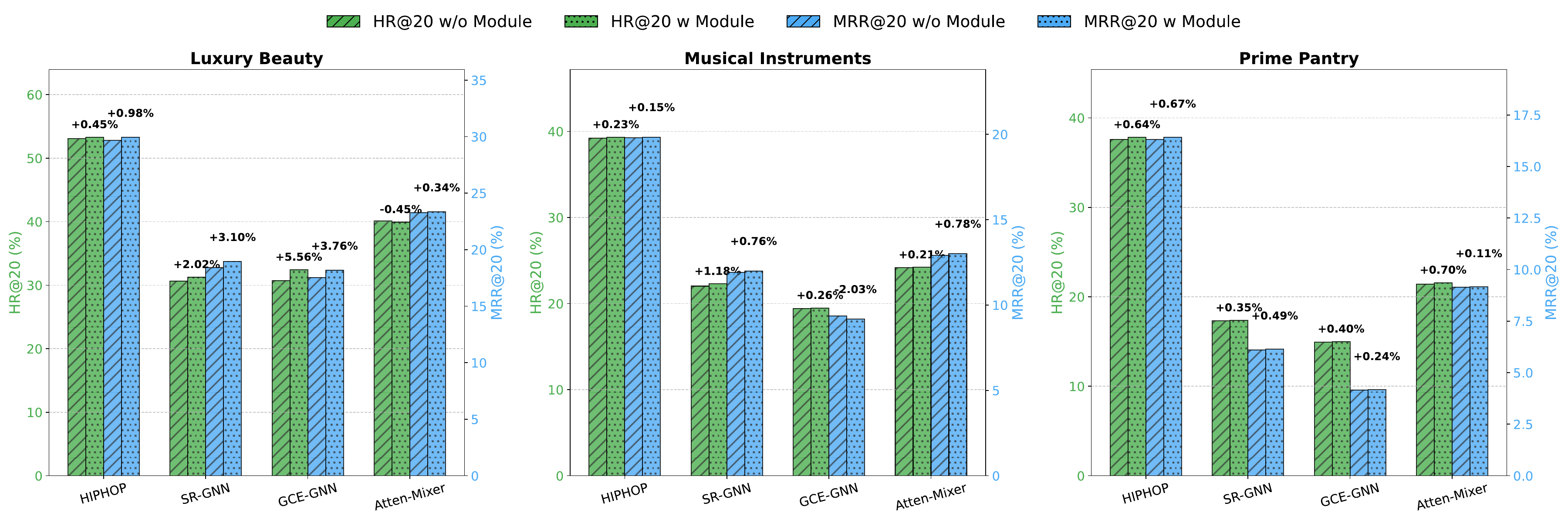}
    \caption{Impact of Pluggable LLM-Driven Semantic Embedding Module on Recommendation Performance.}
    \label{fig:LLM_embedding_impact}
\end{figure*}

\subsection{Ablation Study (RQ2)}\label{sec:RQ2}
To validate the effectiveness of each module in HIPHOP, we conducted ablation studies on the Diginetica and Yoochoose 1/64 datasets, comparing HIPHOP with five variants: \textbf{w/o MultiIntent}, \textbf{w/o InterSim}, \textbf{w/o GlobalSim}, \textbf{w/o LocalSim}, and \textbf{w/o Contrastive}. The results are shown in Figure~\ref{fig:ablation_study_results}.

\begin{enumerate}
    \item \textbf{w/o MultiIntent}: Replacing the dynamic multi-intent capture module with average pooling leads to a slight performance drop, indicating its importance in capturing and aggregating multiple user intents for better performance.
    
    \item \textbf{w/o InterSim}: Removing the inter-session similarity learning module, which captures both long-term and short-term interest, leads to the most significant performance decline, highlighting its critical role in enhancing recommendation relevance by leveraging global and local session similarities.

    \item \textbf{w/o GlobalSim}: Excluding global session similarity results in a moderate decrease, emphasizing the importance of global context in improving recommendation quality.

    \item \textbf{w/o LocalSim}: Removing local session similarity results in a performance drop, illustrating its contribution to refining intent representations.

    \item \textbf{w/o Contrastive}: Without the contrastive learning module, which improves embedding quality by maximizing positive similarity and minimizing negative similarity, performance decreases, thereby confirming its crucial role in improving item representation learning.
\end{enumerate}

\subsection{Hyperparameters Study (RQ3)}\label{sec:RQ3}

\subsubsection{Number of Intent Query Vectors \( M \)}
This hyperparameter in the dynamic multi-intent capture module was tested with values in \{2, 4, 5, 10\}. Increasing \( M \) from 2 to 4 significantly improved HR@20 and MRR@20 on both two public datasets, indicating that more intent query vectors enhance the model's ability to identify and aggregate multiple user intents. However, setting \( M \) beyond 4 led to performance plateauing or slight declines, likely due to overfitting caused by the increased model complexity.

\subsubsection{Number of Recent Items \( k \)}
This hyperparameter in the local session similarity graph was tested with values in \{1, 2, 3, 4, 5\}. Performance improved as \( k \) increased from 1 to 3, demonstrating that considering more recent items in other sessions effectively captures short-term user behavior and enhances local similarity. However, beyond \( k=3 \), performance started to decline, likely due to the introduction of noise from less relevant recent items.

\subsubsection{Number of Negative Samples \( N_{\text{neg}} \)}
This hyperparameter in contrastive learning was tested in \{4, 8, 16, 32, 64, 128, 256\}. On the Diginetica dataset, performance improved as \( N_{\text{neg}} \) increased from 4 to 8, while on Yoochoose 1/64, optimal performance was achieved at \( N_{\text{neg}} = 16 \). However, increasing \( N_{\text{neg}} \) beyond these points led to performance degradation due to the introduction of noise and increased computational overhead.

\subsubsection{Contrastive Loss Weight \( \lambda \)}
The weight coefficient \( \lambda \) for the contrastive loss was tested in \{0.0, 0.01, 0.05, 0.1, 0.3, 0.5, 0.8\}. The optimal performance was observed at \( \lambda = 0.3 \) for Diginetica and \( \lambda = 0.3 \) or \( 0.5 \) for Yoochoose 1/64, effectively balancing the prediction and contrastive loss. However, excessively high values of \( \lambda \) reduced the effectiveness of the prediction task, leading to a decline.

\subsection{Impact of Semantic Embedding (RQ4)}\label{sec:RQ4}

To evaluate the impact of the LLM-driven semantic embedding module on HIPHOP's performance, we conducted experiments on the Luxury Beauty, Musical Instruments, and Prime Pantry datasets. The results, shown in Figure~\ref{fig:LLM_embedding_impact}, reveal a performance decline when the module is removed. For instance, on Luxury Beauty, HIPHOP w achieved HR@20 and MRR@20 scores of 53.30\% and 29.95\%, compared to 53.06\% and 29.66\% for HIPHOP w/o, with similar trends across other datasets.

We also tested the module's portability by integrating it into SR-GNN, GCE-GNN, and Atten-Mixer, showing performance improvements across most models and datasets. For example, SR-GNN's HR@20 increased from 30.65\% to 31.27\% on Luxury Beauty, GCE-GNN's from 30.73\% to 32.44\%, and Atten-Mixer showed enhancements on other datasets. These results confirm the module's effectiveness in boosting performance.

\section{Conclusion}

This paper presents HIPHOP, an SBR model that improves item semantics, session dependencies, and user interest modeling. HIPHOP leverages LLMs for item embeddings, GNNs for item transitions, and a dynamic multi-intent module for complex user interests. Techniques like intent-guided denoising, hierarchical session similarity learning, and contrastive learning improve accuracy and robustness. Experiments show HIPHOP outperforms state-of-the-art methods. Future work will focus on handling complex user behaviors and incorporating multimodal data for cold-start and dynamic scenarios. 



\begin{acks}
This work was supported in part by BNSF(L233034), Fundamental Research Funds for the Central Universities, JLU(93K172024K17, 17221062401) Open Project of Anhui Provincial Key Laboratory of Multimodal Cognitive Computation, Anhui University(MMC202408), GBABRF‌‌(2025A1515010739), GSTP(2024A04j6317) and NSFC(62172443).
\end{acks}

\bibliographystyle{ACM-Reference-Format}
\balance
\bibliography{sample-base.bbl}


\end{document}